\documentclass{ws-p8-50x6-00}

\begin{document}

\title{Numerical Relativity: Status and Prospects}

\author{Luis Lehner\thanks{CITA National Fellow}}

\address{Department of Physics and Astronomy, The University of British Columbia \\
\& Pacific Institute for the Mathematical Sciences, \\
Vancouver, BC, V6T 1Z1, Canada\\ 
E-mail: luisl@physics.ubc.ca}


\maketitle

\abstracts{
We are entering an era where the numerical construction of generic spacetimes
 is becoming a reality.  The use of computer simulations, in principle, allows
 us to solve Einstein equations in their full generality and unravel important
 messages so far hidden in the theory.  Despite the problems still found in present
 applications, significant progress is being achieved in a number of areas where
 simulations are on the verge of providing useful physical information.  
This article reviews recent progress made in the field, paying particular
 attention to black hole spacetimes and discuss open problems and prospects for the future.}

\section{Introduction}
From its beginning a few decades ago,  Numerical Relativity has
enticed researchers with the promise to finally unravel what General Relativity
has kept hidden for almost a century. However, as any other discipline in its
infancy, a costly price has been paid to understand how to  construct spacetimes numerically
and decipher its physical content. Despite significant results obtained in a number
of areas, like the discovery of {\it critical phenomena} in G.R.\cite{mattFIRSTCRITICAL},
some criticisms to the field exist due to progress not being as rapid as some expected.

Although it is certainly true that the pace has not been as fast as hoped, progress has
been steady and the current status of the discipline gives confidence for even the near future.
Just as much has been learned from the theory from analytical (`pencil and paper')
techniques in its first century,  the second century promises to considerably advance our knowledge due 
to our ability to numerically probe interesting scenarios. In this article, I describe the current status
 of the field in a number of fronts and highlight the progress achieved during the past three 
years (from the previous G.R. meeting\cite{seidelGR15,choptuikGR15}). It will be impossible to cover all areas in this
contribution, thus I will concentrate on black hole simulations since it is precisely in 
strongly gravitating/highly dynamical scenarios where this field has its {\it `niche'}. Also, for the 
most part, I will avoid non-vacuum 
scenarios as these will be discussed by Shibata in the present volume.

\section{The Arena}
Achieving a numerical implementation of a PDE system is a complicated task as one tries
to model a `continuum' with a discrete representation. Of all systems to be modeled,
arguably, the most difficult one is given by Einstein equations. Lack of a preferred frame;
gauge issues; a system of ten  quasi-linear tightly coupled equations; singularities; variables 
``being'' the spacetime rather than living `on top of it'; etc. complicate the implementation task.
In fact, even the equations are not a priori expressed in a convenient form for a numerical evolution; i.e.
$E_{ab} \equiv G_{ab} - 8\pi T_{ab}= 0$ does not have an explicit initial value problem in general (IVP). That is, given 
the initial state of the system at a given ``time'', the solution to the future is obtained by integrating 
the equations. The seminal work of Choquet-Bruhat\cite{choquebruhat50} 50 years ago showed the path to follow
to achieve a convenient treatment and since then, many works have extended this and presented different variations
 on how to define an IVP for Einstein equations. 

At present, there are roughly three main `venues' or approaches pursued in numerical relativity to 
cast the problem in a convenient way: the {\it Cauchy}, {\it characteristic}
and {\it conformal approaches}. Several important milestones have been achieved in each of these
approaches and I would present them in separate sections, stressing where each one stood just few
years ago and where they are now which illustrate they have come a considerable way. 
Yet, there are still many hurdles left to be crossed for a robust implementation and where appropriate
I will include some points that would need to be addressed for this purpose.

\subsection{Singularity handling}
Of all the aforementioned problems, a crucial one is handling the singularity which 
can be already present on the initial hypersurface or might be formed at a later time by gravitational
collapse. At the analytical level, a singularity marks the demise of the theory; but at the numerical one, 
even points away from it but ``close'' enough will have fields whose gradients will likely be too large for the implementation.
Thus, if at all possible, one would like to stay away from the singularity and neighboring points. For years,
the prevalent strategy was to use the freedom in choosing the foliation of the spacetime, so as to not evolve the
equations in that region\cite{piranHOUCHES}. By `piling-up' the slices so that the proper-time along the normal 
direction is
zero, one `avoids' the singularity. Although this strategy works for short time evolutions, it induces large
gradients in long evolutions (due to the slices being considerably `bent'). Recently, the prevalent strategy
has become the one known as {\it singularity excision}\cite{unruhexcise}; here the idea is to simply cut-short the evolution
inside the event horizon. For this strategy to be physically valid, two conditions are required: 
First, the validity of cosmic censorship and second a method
to estimate the location of the event horizon with only partial information of the spacetime. The second one is addressed
by locating apparent horizons (with the caveat that they might not exist even where there is an 
event horizon\cite{waldNOAPHOR}) while the first one is still an assumption believed to hold in `reasonable' spacetimes.

\subsection{The Formulations}
As mentioned, there are many ways to cast Einstein equations in convenient form for numerical purposes. These can be
roughly distinguished by the level set of hypersurfaces adopted to foliate the spacetime and whether this
 is the physical one or a `larger' one where the physical description is obtained
a {\it posteriori} by a rather straightforward restriction. A notion of ``time'' is defined by a chosen foliation
of the spacetime (physical or not) which is parametrized by a parameter $\lambda$. The character of the normal to 
the $\lambda=const$ surfaces ($n_a = \nabla_a \lambda$) 
characterizes the approach. A Cauchy/characteristic approach corresponds to  $n_a$ being timelike/null\footnote{Of course
one can deal with more generic conditions but this has not been pursued agressibly yet.}.

In all approaches, the evolution equations are obtained by projecting Einstein equations onto the hypersurfaces.
The projections with respect to the normal (either one or both components of $E_{ab}$) correspond to 
constraint equations which are preserved by the evolution equations if they are satisfied at the initial
hypersurface by virtue of the Bianchi identities. Note however, that this allows one to arbitrarily modify
the evolution equations by adding the constraints to them. Although this defines a totally equivalent
system, the behavior of the numerical implementation can be drastically different as we will discuss
later.

Just as is customary in analytical investigations of the theory, there is far from a preferred approach; we have long
learned that in General Relativity {\it ``one size does not fit all''}. Indeed, different approaches are naturally
suited for different problems and I will describe the main focus of these in current efforts. \\

\noindent {\bf Cauchy approach:}
As mentioned, in this case the spacetime is foliated with spacelike hypersurfaces, as a result, all
$\lambda=const$ surfaces `end' at $i_o$. This brings in a problem as only in a limited number of cases, the
numerical implementation can cover the whole hypersurface due to limited computational 
power\footnote{One can compactify
the spacetime and do include infinity in the numerical grid, however, special care must be taken
as the effective discretization length grows to infinity which can easily render the numerical
implementation useless.}. Hence the hypersurfaces are terminated at an outer boundary defining a finite
domain. The most widely known formulation of Einstein equations within this approach
is the so-called ADM system\cite{mtw}; however, difficulties observed in the implementation of this system
have spurred substantial efforts to consider different reformulations that might aid in the implementation 
(and in several cases there is evidence for this being the case\cite{baumgarteBOUND,golmEVOL}). Aside from 
having spacelike leaves, the choice of foliation
is largely arbitrary; as a result this approach is quite flexible and is certainly the most popular one in
numerical relativity.\\

\noindent {\bf Characteristic approach:}
In this formulation, the hypersurfaces are characteristics of the spacetime. In the particular case
of an isolated system, these hypersurfaces can be chosen to reach future null infinity. Since gravitational
waves propagate along these hypersurfaces the phase of the waves approaches a constant in the far regions. 
The spacetime can be compactified (as suggested by Penrose\cite{penrose} to study asymptotic properties) without
numerical problems and enabling the rigorous evaluation of fields at future null infinity. 
The applicability of this approach is basically restricted to cases where null hypersurfaces do not cross\footnote{There
exist proposals to go around this problem in a variety of cases\cite{friedrich-stewart,c2m} though they might be 
difficult to implement numerically.}
 (i.e. free
of caustics or crossovers), otherwise the coordinate system becomes singular. In caustic-free regions, this approach
is quite useful and simplifies the description when compared to the Cauchy approach as much of the gauge freedom
is fixed and the variables are intimately related to physical quantities\cite{bondi,sachs,tamb-win}.\\

\noindent {\bf Conformal approach:}
This formulation is in a way the most general one as it does not assume any particular condition on the
hypersurfaces, it is thus amenable for a Cauchy, characteristic or more generic implementations. The peculiarity
of this formulation is that it first obtains the description of an unphysical spacetime and the physical one
is obtained a posteriori\cite{friedrichORIG}. The computational boundaries are therefore outside the
physical spacetime, and hence are causally disconnected from it assuming all leaves of the foliation cross future
null infinity.

\section{Status}
\subsection{Cauchy approach \label{sec:Cauchy}}
\noindent {\bf Recent Past:}
A few years ago numerical simulations within this approach presented
a series of remarkable results. Among these, the discovery of critical
phenomena in GR in spherical symmetry (1D)\cite{mattFIRSTCRITICAL,carstenreview}, axisymmetric (2D)
head-on  collisions of black holes\cite{seidelHEADON}
and collapse of Vlasov systems\cite{shapiroteukolskyTORUS}. 1D
simulations were implemented with impressive accuracy thanks to the use
of adaptive mesh refinement (AMR), [see for instance\cite{bergeroliger}]. 2D simulations on the other hand, barely had
enough resolution to resolve the essential features and since singularity excision was used
they could not proceed too much into the future. The need to produce generic simulations capable
of producing waveforms templates for gravitational wave data analysis shifted the attention onto
three-dimensional (3D) implementations. In a way, 
this was a bit regrettable as many interesting physical systems which are truly (or basically) two 
dimensional in nature were left aside.
Currently, activities in both 2D and 3D are aggressively carried out, a few examples are:\\

\noindent {\bf 2D:}
Recently, a few groups have undertaken the effort to produce 2D 
simulations with the necessary numerical techniques to obtain the
desired accuracy to address problems like critical collapse in axisymmetry; head-on
collisions of black holes and accretion onto a black hole. Among these AMR and singularity
excision are the most notable ingredients. The implementation in~\cite{mattGRAXI} is concentrating on studying critical
phenomena in Einstein-Klein-Gordon systems to investigate critical behavior in more generic
settings. The ADM formulation is being used with a partially constrained implementation
(i.e. some of the constraints are used to update the field variables) which apparently helps to remove exponential
growing modes typical of free evolution schemes (where the constraints are just used to monitor the quality of
the solution) and encouraging results are being obtained. For instance, simulations describing two separated
`pulses' of scalar field in a slightly super-critical configuration are followed. As time progresses, each
pulse collapses forming two separated black holes which later
merge into one which then settles into an almost stationary black hole. At late times the accuracy deteriorates but this
will likely be solved once AMR gets fully incorporated. \\

\noindent {\bf 3D:}
Three dimensional simulations still have a number of  unresolved issues. Poor 
resolution; lack of understanding of `ideal formulations'; boundary treatment and
initial data definitions are among the outstanding problems being studied at present.
Despite these difficulties, significant advances have been presented recently. Next I comment on
some of these,\\

\noindent{\it Initial Data:}
On the initial data problem, work is advancing towards obtaining realistic
initial data. Within this approach, the problem of posing consistent initial data 
reduces to providing the intrinsic metric of the initial hypersurface and its extrinsic curvature
satisfying the constraints (Hamiltonian and momentum constraints). A procedure to handle this
problem was developed by Lichnerowicz\cite{lichnerowiczID} and further refined by York\cite{yorkID} which
involves specifying a conformally related `seed' metric (see \cite{cookID} for a detailed description).

For decades the prevalent approach was to solve this Lichnerowicz-York\cite{yorkID}
problem assuming a conformally flat seed metric (supplemented by an appropriate seed for a quantity related
to the extrinsic curvature).
However, since the Kerr-Newman spacetime can not be obtained this way\cite{priceCONFORMALFLAT}, spinning black holes which
 can be obtained under this assumption must contain
spurious gravitational radiation. A series of proposals have been presented to give away with
conformal flatness\cite{jeffID,matznerID,pedroID_I,dainID} and implementations of these are 
already in place or being developed\cite{pedroID_II,dienerAMR,teukolskyAGAINSTCONF}.
Reasonable arguments indicate that these initial should data contain less spurious radiation,
but it is yet unclear how much there is. This issue will be fully resolved by numerical
evolutions. A key ingredient that needs to be worked out for treating binary systems is
to choose the initial data achieving a smooth transition from the Post-Newtonian phase
description of the inspiraling phase. This will guide the choice of `seed variables' in a way consistent with the 
description of the problem and aid in minimizing spurious radiation. Work in this area is just beginning.\\

\begin{figure}[t]
\begin{center}
\epsfxsize=27pc 
\epsfbox{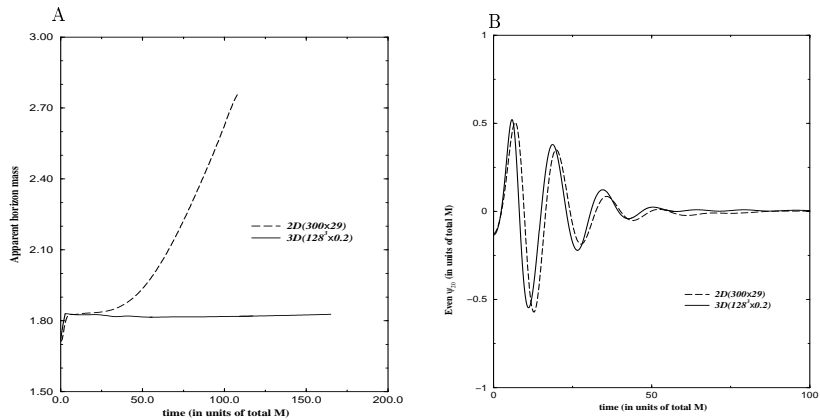} 
\end{center}
\caption{Plots of: (A) the apparent horizon area obtained with the 2D (dashed line)
and 3D codes (solid line) vs. time, the problems induced by the use of singularity avoidance techniques are
apparent in the growth of the apparent horizon which is not present in the 3D code's result. 
(B) Gravitational wave extraction with both codes, for the length of simulations considered with the 2D codes,
the results are in good agreement. \label{fig:golm}}
\end{figure}

\noindent{\it Single Back Hole evolutions:}
Singularity excision has also allowed for extending simulations of single black hole spacetimes
 by several of orders of magnitude when compared to similar spacetimes modeled with 
singularity avoiding techniques (see figure \ref{fig:golm}). At least for a number of cases, there are now simulations 
that can 
evolve single perturbed black holes for considerable lengths of time, certainly longer than needed for studying
a variety of scenarios; probing regimes beyond what we know from perturbation theory and 
study the produced waveforms. The work presented in\cite{golm}, illustrates how singularity
excision, coupled with a carefully designed coordinate condition that minimizes changes in particularly
sensitive variables in the BSSN\footnote{BSSN is a variant of the ADM 
formulation which has become quite popular due to its better performance in several situations when compared to
the ADM one.} formulation\cite{shibnakam,baumgarteBOUND}, enables practically unlimited evolutions of spacetimes
describing black holes perturbed by `Brill waves'. Not only do these results outperform
the very best ones obtained with `old' 2D codes (with singularity avoidance), but it is used to accurately extract
the expected waveforms. \\

\begin{figure}[t]
\epsfxsize=26pc 
\epsfbox{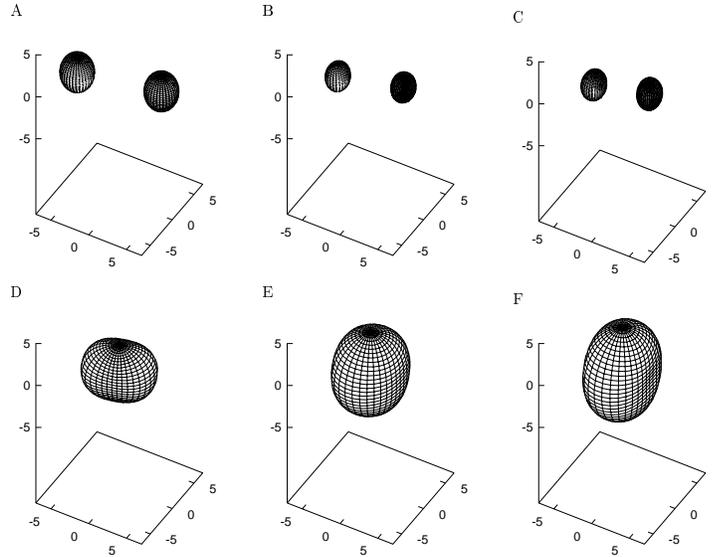} 
\caption{Snaphots of apparent horizon locations during the evolution (A-F corresponding
to $t=0,2M,4M,6M,10M,12M$). At $t=0$ (A) two distinct
horizons, separated by $~10M$, at $t~5M$ a single one is already formed which acquires a more
spherical shape as time proceeds (and the evolution is still well behaved). \label{fig:2bh}}
\end{figure}

\noindent {\it Binary Back Hole evolutions:}
Binary black hole systems are among the `hottest' problems for numerical relativity. This system is a prime
candidate for producing gravitational radiation to be measured by the new generation of gravitational wave
detectors. An accurate description of the system requires numerical simulations of the full Einstein
equations especially for the merger phase. In dealing with such a problem numerically, one must handle two
initially separated black holes which will merge and settle into a single perturbed black hole (which can then
be treated by black hole perturbations\cite{pricepullinCLAP,lazarus}). The total length of simulations required is 
at least in the
order of hundreds of M's (from about the inner most stable circular orbit to the point where perturbations around
a single black hole can be safely used). There is still no implementation capable of addressing these requirements
but some hopeful preliminary investigations have been presented. In particular, a
 proof-of-principle implementation was presented demonstrating the ability to simulate the system
 containing two initially separated singularities to past the merger phase with the 
use of singularity excision.  The results in\cite{bhcollision2} illustrated an evolution of a system whose 
initial data described two-separated (by about $10M$) equal mass ($M$)
black holes (judging from the locations of the apparent horizons)  and followed the dynamics of the system to well
past the merger.  Figure \ref{fig:2bh} illustrates the location of the apparent horizons at different times. 
Although the simulation crashes after a few  tens of Ms, this work showed that 
singularity excision techniques can handle the binary black hole problem. \\

\noindent {\it Towards the attainment of the `ultimate' reformulation:}
Considerable efforts are being spent to understand why different formulations of Einstein equations
yield simulations behaving quite different. The difficulties observed in numerical implementations 
spurred a large number of works that presented re-formulations of Einstein equations in explicit symmetric
hyperbolic form (see \cite{reulaREVIEW} for a detailed list). These formulations write the system in first order
form (and hence require introducing additional variables to express spatial derivatives) and manipulate
the equations through the addition of constraints to guarantee the existence of a complete set of eigenvectors (of
the matrices describing the the principal part of the system).
Several benefits are gained after all this work\footnote{See Rendall's contribution to this volume.}: the standard mathematical machinery for PDE's can be used to determine 
the well posedness of the problem under study;
whether the characteristic speeds of the system are physical (lie inside the null cones) and furthermore, determine which
combination of variables are ingoing and outgoing with respect to a given boundary (which plays an important role when
 imposing boundary conditions). In a well posed problem, the solution at a later time, $S(t)$, satisfies
$|S(t)| \leq \alpha e^{\kappa t} |S(t=0)|$ (with $\alpha, \kappa \in {\bf R}$). This property guarantees continuous
dependence on the initial data which is important as the initial data provided in a numerical implementation is, at best,
the desired one only up to round-off errors.  The fact that the solution is bounded by an exponential function
though, is not sufficient information for the numerical implementation, as one would like to rule out exponential
modes in some given problems. This is the case, for instance, when modeling a black hole spacetime in an appropriately chosen gauge
which, even when perturbed, should relax to a stationary black hole.

\begin{figure}[t]
\begin{center}
\epsfxsize=20pc 
\epsfysize=16pc
\epsfbox{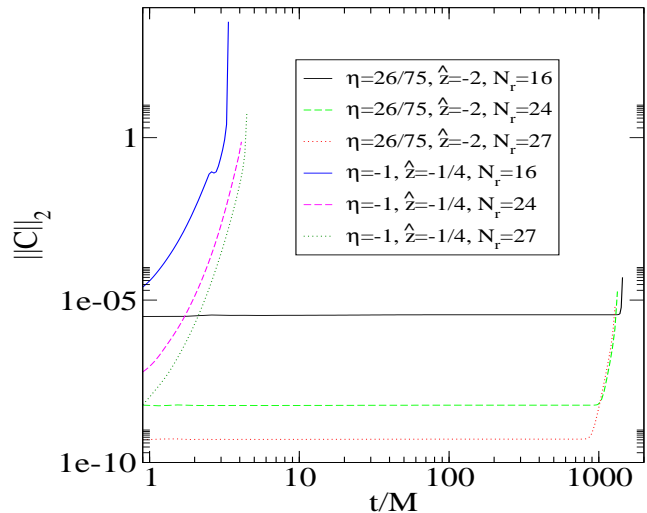} 
\end{center}
\caption{Solution behavior for two members of the family of formulations (parametrized by
$\eta$, $\hat z$). For the same resolutions (given by $N_r$) the results
of both differ considerably.\label{fig:cornell}}
\end{figure}

Clearly, given any formulation one can write infinitely many others by simply adding constraints to the right hand
sides with free multiplicative parameters. Although at the analytical level all these
formulations are exactly equivalent; at the numerical one they might display considerable differences.
Examples of such behavior have been illustrated in 3D~\cite{generalEC} and in 1D~\cite{pabloK}. For
 instance, in~\cite{generalEC} a two-parameter hyperbolic family of reformulations of Einstein equations 
has been presented which have {\it exactly} the same principal part. Therefore, well posedness of systems treated by
all members of
the family  follows in a totally equivalent manner. This work presented results obtained by simulating Schwarzschild
spacetimes with different members of this family under the {\it same} numerical implementation. Total evolution
times ranged from as short as $6M$ to $1500M$. Figure \ref{fig:cornell} illustrates the norm of a representative constraint
vs. time for two different members of the family of formulations (under the same discretization and resolution). While 
for one, the violations become very severe at earlier times, for the other only at quite late times do these become
apparent. The message from these and other works is that for numerical simulations, 
{\it well posedness is not enough!} There is an urgent need for a method to discern from the ``zoo'' of formulations
those that might simplify the attainment of a stable implementation. Formulating such a method is certainly 
non-trivial, as one
would need to be able to separate ``gauge modes'' from physical ones; understand the behavior of constraints
off the constraint surface and the influence of non-principal part terms in the evolution as well as that
of boundaries. Nevertheless, its impact on the field would likely be profound.\\

\noindent {\it Boundary conditions:}
Outer boundary conditions are difficult to define in any theory containing propagating waves. It
 is particularly
difficult in relativity due to backscattering, gauge and constraint issues. Approximate boundary 
conditions have
been used in the past with different degrees of effectiveness (see~\cite{luisreview} for a list). However, not 
only is it not clear how much
spurious reflections or incoming waves these introduced but most importantly whether are these
even consistent. At the core of this question lies the fact that `standard' boundary conditions
provide values to all variables and whether these conditions satisfy the constraints.
Recently, insight into this problem and its physical interpretation has been presented
in\cite{friedrichnagy} where a careful analysis was presented that studied the question of well posedness
of the problem considering the presence of boundaries. The conclusion is that only {\it two} pieces of information
can be given (related to the  polarizations of incoming radiation) while the rest are determined 
by the Einstein equations themselves. 

The need for having correct boundary conditions
respond to several reasons: (i) Placing boundaries as `close as possible' to minimize computational
costs; (ii) Eliminate constraint violating mode sources, (iii) Eliminate spurious reflections . Recently a 
number of efforts (in linearized gravity in 3D\cite{jeffBOUND} or full gravity in spherical symmetry\cite{chicholuisCON}) 
have shown that carefully treated boundaries can satisfy these requirements benefiting
the numerical evolution.

\subsection{Characteristic approach}
\noindent {\bf Recent Past:}
Numerical implementations within this formulation have come a long way despite
it being `newer' than the Cauchy ones. In the recent past the characteristic
formulation has been demonstrated to be a useful tool by producing the first 
unlimited evolution of single black hole spacetimes\cite{movebh}. Work within this formulation
is concentrating on scenarios or spacetime regions which are expected to be
free of caustics. I will here mention a few examples both in the vacuum and non-vacuum
cases (I include the latter since it is not being covered by Shibata's contribution to this volume).

Unfortunately, because of caustic formation, this formulation can not address 
all problems of interest, in particular binary black holes or binary neutron stars.
However for certain systems back of the envelope calculations indicate they should be free of this problem .

Motivated by these problems a series of works were devoted to understand the incorporation
of matter within the formulation. Most notably the implementation of high resolution schemes\cite{philipMATTER}
and the simulation of weak pressure fluids in full 3D black hole spacetimes\cite{mattchar}.\\

\noindent {\bf Present focus}\\

\begin{figure}[t]
\begin{center}
\epsfxsize=19pc 
\epsfysize=15pc 
\epsfbox{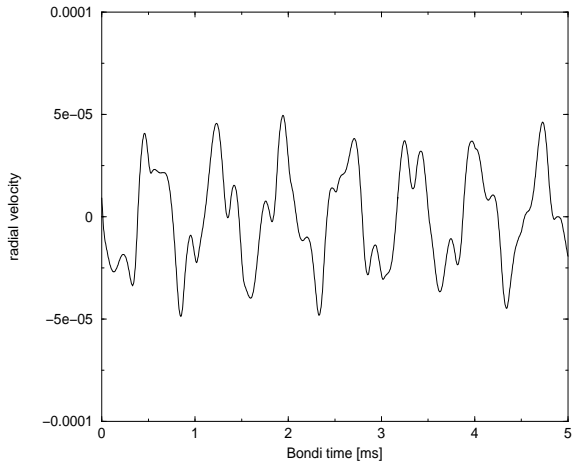} 
\end{center}
\caption{Evolution of the radial component of the star's velocity.
\label{fig:papa}}
\end{figure}

\noindent {\bf 2D:}
Much activity is being carried out with $2D$ characteristic codes\cite{dinvernoEVOL,philipMATTER}. In particular,
studies of neutron star perturbations in full general relativity are being
carried out where the dynamics of the fluid is described by high resolution
shock capturing schemes\cite{philipMATTER}. In these works, the equilibrium configuration
of the star is perturbed and additional incoming
gravitational waves are considered in some tests.  The oscillations of the star are accurately modeled and compared
with perturbative results and the gravitational radiation is read-off at future null infinity. Figure \ref{fig:papa}
illustrates the oscillations a polytropic star which has been perturbed
by slight density and pressure deviations from the equilibrium configuration.
Mass conservation is verified to an impressive degree of accuracy which coupled to a series of
other successful tests indicate the code is ready to tackle interesting scenarios. \\

\noindent {\bf 3D:}
Non-vacuum 3D simulations within this approach are targeting spacetimes containing
a black hole and a compact star. Proof of principle tests to examine the feasibility of
extending the vacuum code described in\cite{hpgn,cce} to the non-vacuum case proved successful
in very weak pressure cases\cite{mattchar}. Efforts are now concentrated in providing correct
initial data describing a star ``close'' to the black hole. The main reason for this is to avoid
caustics. A back of the envelope calculating using the lens equation implies that the
spacetime will be free of caustics provided $D < R_s^2/(4m_s)$ (where $R_s, m_s$ are the are radius
and mass of the star which is at a distance $D$ of a unit mass black hole). This defines a window
of applicability which does contain interesting sources for gravitational wave detectors. Since neutron star-black hole
systems are also likely sources for gravitational wave detection, simulations of this systems will have 
immediate practical uses.

\begin{figure}[t]
\begin{center}
\epsfxsize=24pc 
\epsfbox{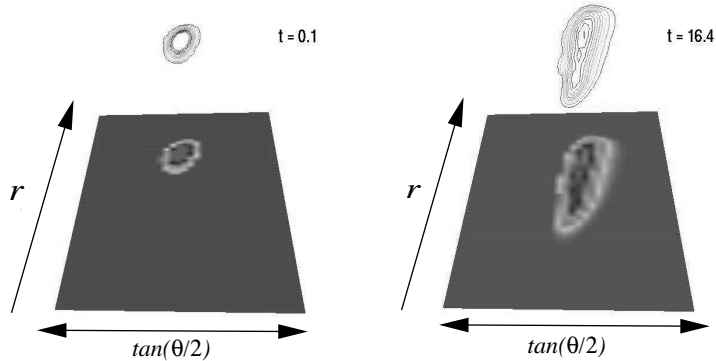} 
\end{center}
\caption{Two density snapshots in the at $\phi=const$, as time progresses the tidal distortion
of the polytrope becomes evident. \label{fig:rho}}
\end{figure}

The implementation to treat this problem
follows closely the $2D$ one as far as treatment of the fluid equations is concerned but,
being still a serial code, it can not achieve as high resolutions in full 3D and some numerical problems are
still being addressed. Once the level of resolution needed is achieved, the simulation should perform
as well as the one in the 2D case. Figure \ref{fig:rho} shows a $\Gamma=2$ polytrope with radius $5M$ 
which is placed at a distance of $9M$ from a mass $M$ black hole and moves in the $\theta$ direction.
Since a portion of the star lies inside ISCO, as time progresses it becomes quite distorted. Results obtained
with this simulation are qualitatively correct but they still have inadequate accuracy for producing useful gravitational
wave templates.

On the vacuum front, a project aimed towards
obtaining gravitational radiation of a binary black hole spacetime is also under development\cite{jeffPROGRESS}.
Here, the spacetime is envisioned in a time-reversed point of view. This is motivated by
the possibility of posing a double null problem whose inner boundary corresponds to
a fissioning white hole\cite{fissionwh,husaTORUS} (which in a time reversed point of view corresponds
to merging black holes)
and the other corresponds to ${\cal I}^-$. An inverse scattering process can be formulated
to obtain the radiation produced by a binary black hole collision\cite{gomezFISSION}. Preliminary investigations
 of this approach
have targeted a ``close limit approximation'' yielding excellent results\cite{jeffFISSIONCLOSE}. Of course,
since the theory is non-linear this procedure is far from direct. However,  note the data needed for this problem
is intimately related to the radiation fields at both scri's which are expected to be quite small
(in fact should be `zero' at ${\cal I}^-$ and about $5\%$ of the total mass at ${\cal I}^+$); thus, one can formulate (partially
hand-waving) arguments to give support to this approach in terms of an iterative procedure.

\subsection{Conformal approach}
As mentioned within this approach, the description of an unphysical spacetime is obtained and
a restriction of the solution yields the physical spacetime (and the metric of the latter is
conformally related to that of the former). All implementations within this 
approach have used a Cauchy initial value problem formulation thus have many similarities to what
was included section\ref{sec:Cauchy}. There is an extra gauge issue here since the relation between
the conformal metrics is essentially arbitrary. \\

\noindent {\bf Recent Past:}
The conformal implementation of Einstein equations has proceeded at a steady pace since
it was first used to investigate critical phenomena in 1D\cite{peter1D} and subsequently extended to
higher dimensional spacetimes\cite{frauendienerEVOL,peterITP}. \\

\noindent {\bf Present:}
Preliminary proof-of-principle simulations of Schwarzschild black holes in this formulation have
been presented in~\cite{peterITP}. These targeted the Kruskal extension and simulations proceeded
well for moderate amounts of times.

\begin{figure}[t]
\begin{center}
\epsfysize=14pc 
\epsfxsize=20pc
\epsfbox{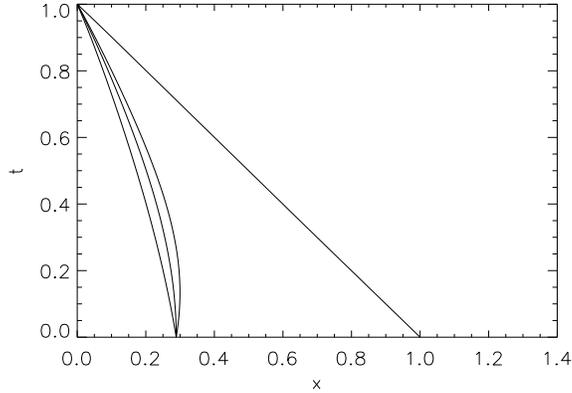} 
\end{center}
\caption{Timelike geodesics in a linearly perturbed flat spacetime and ${\cal I}^+$ meet at ${\it i}^+$.\label{fig:infinity}}
\end{figure}

More recently, linear waves on Minkowski spacetimes have been successfully evolved
reconstructing the spacetime all the way to ${\it i}^+$~\cite{husa:2001ad}. Since  ${\it i}^+$ is safely inside
the computational domain, this formulation allows for studying the regularity and structure
of the whole spacetime. Figure \ref{fig:infinity} illustrates three different timelike geodesics in flat spacetime
perturbed linearly. As expected, all end at the same point in the future which meets ${\cal I}^+$, it is confirmed
that ${\it i}^+$ is regular by studying the spacetime structure in its neighborhood.
Although more experience still needs to be gained to exploit this formulation
(particularly in gauge related issues), it is becoming clear that this formulation is indeed a useful tool
in numerical relativity and the ideal one to study global properties of spacetimes. \\

\section{Future Problems}
It is certainly true that most current efforts in the field are directed towards
systems which will likely produce gravitational waves detectable by the new generation
of detectors. However, this is just the tip of the iceberg. The ability to solve
Einstein equations in generic scenarios will open the door for studying a variety of
problems, a few examples being
\begin{itemize}
\item{{\it Astrophysically relevant systems:}}
A series of spectacular and yet poorly understood phenomena ``illuminate'' the sky
with powerful energetic emissions. Active galactic nuclei, gamma ray bursts, quasars,
micro-quasars, give away so much energy in relatively short  periods of time (but not so 
short that ``explosive'' mechanisms can be called into place; see Piran's contribution 
to this volume) that it is believed that a compact central object, a fairly massive surrounding
plasma and magnetic fields are key ingredients of the central engine powering these
emissions. Modeling such systems
accurately require solving Einstein equations in the highly dynamical/strong field scenarios.
Future research in this field will certainly have strong emphasis towards astrophysically
motivated problems. 
\item{{\it Spacetimes on the large:}}
Numerical simulations are also opening the door to understand the structure of
spacetimes in the large. Under certain assumptions the asymptotic behavior of spacetimes
is well understood but only in a limited number of cases does one have an explicit
relation between sources and fields at infinity. Numerical implementations can be used
to search for this relation and to examine asymptotic fields behavior without past adopted assumptions
to render the problem analytically tractable.
\item{{\it Quantum Gravity:}}
Another area where simulations might provide valuable insights on prospective quantum 
gravity theories. First attempts in this direction are studying consequences of
the AdS/CFT duality\cite{husainCHARACT}, and the final fate of black strings\cite{blackstrings}
which will provide insights in the classical or low energy limit of possible theories. By
analyzing this limit, one can foresee that unacceptable results will help in ruling out some candidates
while others might help us understand acceptable ones.
\end{itemize}

\section{Final Comments}
As has been evident from the contributions to the numerical relativity sessions in this
meeting and recently published results notable advances are taking place in this field.
 It is certainly impossible to do justice to all efforts in numerical
relativity in this contribution\footnote{For more details please refer to
\cite{luisreview,carstenreview,beverlyREVIEW,cookREVIEW,anninosREVIEW,piotrarticle}.}. I hope this article gives a current glimpse of the activity in
the field and, especially, highlights that considerable progress has been achieved
in the past few years. Undoubtedly, we still face many problems to realize the
full potential of this activity and researchers are working hard towards their
resolutions. Numerical techniques, complementing the analytical ones are paving
the way for understanding the full implications of the theory; certainly, the second century
of General Relativity promises be an exciting one.

\section{Acknowledgements}
I would like to thank N. Bishop, F. Pretorius and F. Siebel for helpful comments on early versions of the manuscript.
I thank M. Alcubierre, S. Husa, M. Scheel and F. Siebel for providing some of the figures included in this text. 
Computations of some of the results presented here were performed on the vn.physics.ubc.ca Beowulf cluster
which was funded by the Canadian Foundation for Innovation (CFI). I want to express my gratitude to the 
California Institute of Technology for its hospitality where the last parts 
of this work were completed.
Finally it is a great pleasure to thank the organizers of GR16 for such an exciting conference.


\end{document}